The paper is written in the ICHEP proceedings style.
The corresponding LaTeX
style file `ichep.sty' is attached right after the `\end{document}'
command of the paper.

--- snip --- snip --- snip --- snip --- snip --- snip --- snip --- snip ---

\documentstyle{ichep}
\def\vcb{\mid V_{cb} \mid}
\def\vtd{\mid V_{td} \mid}
\def\vub{\mid V_{ub}/V_{cb} \mid}

\def\o{\over}
\def\b{\begin{equation}}
\def\e{\end{equation}}

\def\kpnn{K^+\rightarrow\pi^+\nu\bar\nu }
\def\kpn{K^+\rightarrow\pi^+\nu\bar\nu}
\def\klpnn{K_L\rightarrow\pi^0\nu\bar\nu}
\def\klpn{K_L\rightarrow\pi^0\nu\bar\nu}

\begin{document}

 \title{
 {\normalsize\rm
\rightline{MPI-PhT/94-57}
 \rightline{TUM--T31--70/94}
 \rightline{August 1994}
 \ \\
 \ \\
 \ \\}
 Towards Precise Determinations of the CKM \\
  Matrix without Hadronic Uncertainties*}

\author{Andrzej J. Buras}

\affil{ Technische Universit\"at M\"unchen, Physik Department\\
 D-85748 Garching, Germany \\
 Max-Planck-Institut f\"ur Physik
 -- Werner-Heisenberg-Institut --\\
 F\"ohringer Ring 6, D-80805 M\"unchen, Germany}

\abstract{
We illustrate how the measurements of the CP asymmetries in
$B^0_{d,s}$-decays together with a measurement of
$Br(K_L\to \pi^\circ\nu\bar\nu)$ or $Br(\kpnn)$
and the known value of $\mid V_{us}\mid $ can determine all elements
of the Cabibbo-Kobayashi-Maskawa matrix essentially without any hadronic
uncertainties. An analysis using the ratio $x_d/x_s$ of $B_d-\bar B_d$
to $B_s-\bar B_s$ mixings is also presented.}

\twocolumn[\maketitle]

\fnm{7}{Invited talk given at ICHEP '94, Glasgow, July 1994.
 Supported by the German
   Bundesministerium f\"ur Forschung und Technologie under contract
   06 TM 732 and by the CEC science project SC1--CT91--0729.}

\section{Setting the Scene}
An important target of particle physics is the determination of
the unitary $3\times 3$ Cabibbo-Kobayashi-Maskawa
matrix which parametrizes the charged current interactions of
 quarks:
\begin{equation}\label{1j}
J^{cc}_{\mu}=(\bar u,\bar c,\bar t)_L\gamma_{\mu}
\left(\begin{array}{ccc}
V_{ud}&V_{us}&V_{ub}\\
V_{cd}&V_{cs}&V_{cb}\\
V_{td}&V_{ts}&V_{tb}
\end{array}\right)
\left(\begin{array}{c}
d \\ s \\ b
\end{array}\right)_L
\end{equation}
It is customery these days to parametrize these matrix by the four Wolfenstein
parameters  $(\lambda, A, \varrho, \eta)$. In particular
one has
\begin{equation}\label{2.75}
\mid V_{us}\mid=\lambda
\qquad
\vcb=A \lambda^2
\end{equation}
and
\begin{equation}\label{2.75a}
V_{ub}=A\lambda^3 (\varrho-i\eta)
\qquad
V_{td}=A\lambda^3 (1-\bar\varrho-i\bar\eta)
\end{equation}
Here following \cite{BLO}
 we have introduced
\begin{equation}\label{3}
\bar\varrho=\varrho (1-\frac{\lambda^2}{2})
\qquad
\bar\eta=\eta (1-\frac{\lambda^2}{2}).
\end{equation}
which allows to improve the accuracy of the Wolfenstein parametrization.

{}From tree level K decays sensitive to $V_{us}$ and tree level B decays
 sensitive to $V_{cb}$ and $V_{ub}$ we have:
\begin{equation}\label{2}
\lambda=0.2205\pm0.0018
\qquad
\mid V_{cb} \mid=0.039\pm0.004
\end{equation}
\begin{equation}\label{2.94}
R_b \equiv  \sqrt{\bar\varrho^2 +\bar\eta^2}
= (1-\frac{\lambda^2}{2})\frac{1}{\lambda}
\left| \frac{V_{ub}}{V_{cb}} \right|=0.36\pm0.14
\end{equation}
corresponding to
\begin{equation}\label{2a}
\left| \frac{V_{ub}}{V_{cb}} \right|=0.08\pm0.03
\end{equation}
$R_b$ is just the length of one side of the rescaled unitarity triangle
in which the length of the side on the $\bar\varrho$ axis is equal
unity.
The length of the third side is governed by $\vtd$ and is given by
\begin{equation}\label{2.95}
R_t \equiv \sqrt{(1-\bar\varrho)^2 +\bar\eta^2}
=\frac{1}{\lambda} \left| \frac{V_{td}}{V_{cb}} \right|
\end{equation}
In order to find $R_t$ one has to go beyond tree level decays.

As we have seen at this conference a large part in the errors quoted in
(\ref{2}), (\ref{2.94}) and (\ref{2a}) results from theoretical (hadronic)
 uncertainties.
Consequently even if the data from CLEO II improves in the
future, it is difficult to imagine at present that in the tree level B-decays
a better accuracy than $\Delta\vcb=\pm 2\cdot 10^{-3}$ and
$\Delta\vub=\pm 0.01$ ($\Delta R_b=\pm 0.04$) could be achieved unless some
dramatic improvements in the theory will take place.

The question then arises whether it is possible at all to determine the
CKM parameters without any hadronic uncertainties.
The aim of this contribution is to demonstrate that this is indeed possible.
To this end one has to go to the loop induced decays or transitions
governed by short distance
physics. We will see that in this manner clean and
precise determinations of $\vcb$, $\vub$, $\vtd$, $\varrho$ and $\eta$
can be achieved. Since the relevant measurements will take place only
in the next decade, what follows is really a 21st century story.

It is known that many loop induced decays contain also hadronic uncertainties
\cite{AB94A}.
Examples are $B^0-\bar B^0$ mixing, $\varepsilon_K$ and
 $\varepsilon'/\varepsilon$.
Let us in this connection recall the expectations from a "standard" analysis
of the unitarity triangle which is based on $\varepsilon_K$, $x_d$
giving the size of $B^0-\bar B^0$ mixing,
$\vcb$ and $\vub$ with the last two extracted from tree level decays.
As a recent analysis \cite{BLO} shows, even with optimistic assumptions
about the theoretical and experimental errors it will be difficult to
achieve the accuracy better than $\Delta\varrho=\pm 0.15$ and
$\Delta\eta=\pm 0.05$ this way. Therefore in what follows we will
only discuss the four finalists
in the field of weak decays which
essentially are free of hadronic uncertainties.
\section{Finalists}
\subsection{CP-Asymmetries in $B^o$-Decays}
The CP-asymmetry in the decay $B_d^\circ \rightarrow \psi K_S$ allows
 in the standard model
a direct measurement of the angle $\beta$ in the unitarity triangle
without any theoretical uncertainties
\cite{NQ}. Similarly the decay
$B_d^\circ \rightarrow \pi^+ \pi^-$ gives the angle $\alpha$, although
 in this case strategies involving
other channels are necessary in order to remove hadronic
uncertainties related to penguin contributions
\cite{CPASYM}.
The determination of the angle~$\gamma$ from CP asymmetries in neutral
B-decays is more difficult but not impossible
\cite{RF}. Also charged B decays could be useful in
this respect \cite{Wyler}.
We have for instance
\begin{equation}\label{113c}
 A_{CP}(\psi K_S)=-\sin(2\beta) \frac{x_d}{1+x_d^2},
\end{equation}
\begin{equation}\label{113d}
   A_{CP}(\pi^+\pi^-)=-\sin(2\alpha) \frac{x_d}{1+x_d^2}
\end{equation}
where we have neglected QCD penguins in $A_{CP}(\pi^+\pi^-)$.
Since in the usual unitarity triangle  one side is known,
it suffices to measure
two angles to determine the triangle completely. This means that
the measurements of $\sin 2\alpha$ and $\sin 2\beta$ can determine
the parameters $\varrho$ and $\eta$.
The main virtues of this determination are as follows:
\begin{itemize}
\item No hadronic or $\Lambda_{\overline{MS}}$ uncertainties.
\item No dependence on $m_t$ and $V_{cb}$ (or A).
\end{itemize}
\subsection{$K_L\to\pi^o\nu\bar\nu$}
$K_L\to\pi^o\nu\bar\nu$ is the theoretically
cleanest decay in the field of rare K-decays.
$K_L\to\pi^o\nu\bar\nu$ is dominated by short distance loop diagrams
involving
the top quark and proceeds almost entirely through direct CP violation.
The last year calculations
\cite{BB1,BB2} of next-to-leading QCD corrections to this decay
considerably reduced the theoretical uncertainty
due to the choice of the renormalization scales present in the
leading order expression \cite{DDG}.
Typically the uncertainty in $Br(\klpnn$) of $\pm 10\%$ in the
leading order is reduced to $\pm 1\%$.  Since the relevant hadronic matrix
elements of the weak current $\bar {s} \gamma_{\mu} (1- \gamma _{5})d $
can be measured in the leading
decay $K^+ \rightarrow \pi^0 e^+ \nu$, the resulting theoretical
expression for Br( $K_L\to\pi^o\nu\bar\nu$)
is
only a function of the CKM parameters, the QCD scale
 $\Lambda \overline{_{MS}}$
 and  $m_t$.
The long distance contributions to $K_L\to\pi^o\nu\bar\nu$ are negligible.
We have then:
\begin{equation}\label{3c}
Br(\klpnn)=1.50 \cdot 10^{-5}\eta^2\mid V_{cb}\mid^4 x_t^{1.15}
\end{equation}
where $x_t=m_t^2/M_W^2$ with
$m_t\equiv\bar m_t(m_t)$. The main features of this decay are:
\begin{itemize}
\item No hadronic uncertainties
\item  $\Lambda_{\overline{MS}}$ and renormalization scale uncertainties
at most $\pm 1\%$.
\item Strong dependence on $m_t$ and $V_{cb}$ (or A).
\end{itemize}
\subsection{$\kpnn$}
$K^+\to\pi^+\nu\bar\nu$ is CP conserving and receives
contributions from
both internal top and charm exchanges.
The last year calculations
\cite{BB1,BB2,BB3} of next-to-leading QCD corrections to this decay
considerably reduced the theoretical uncertainty
due to the choice of the renormalization scales present in the
leading order expression \cite{DDG}.
Typically the uncertainty in $Br(\kpnn$) of $\pm 20\%$ in the
leading order is reduced to $\pm 5\%$.
The long distance contributions to
$K^+ \rightarrow \pi^+ \nu \bar{\nu}$ have been
considered in \cite{RS} and found to be very small: two to three
orders of magnitude smaller than the short distance contribution
at the level of the branching ratio.
$K^+\to\pi^+\nu\bar\nu$ is then the second best decay in the field
of rare decays. Compared to $\klpnn$ it receives additional uncertainties
due to $m_c$ and the related renormalization scale. Also its QCD scale
dependence is stronger. Explicit expressions can be found in
\cite{BB3,BB4}.
The main features of this decay are:
\begin{itemize}
\item Hadronic uncertainties below $1\%$
\item  $\Lambda_{\overline{MS}}$, $m_c$ and renormalization scales
 uncertainties at most $\pm (5-10)\%$.
\item Strong dependence on $m_t$ and $V_{cb}$ (or A).
\end{itemize}
\subsection{$B^o-\bar B^o$ Mixing}
Measurement of $B^o_d-\bar B^o_d$ mixing parametrized by $x_d$ together
with  $B^o_s-\bar B^o_s$ mixing parametrized by $x_s$ allows to
determine $R_t$:
\begin{equation}\label{107b}
R_t = \frac{1}{\sqrt{R_{ds}}} \sqrt{\frac{x_d}{x_s}} \frac{1}{\lambda}
\end{equation}
with $R_{d,s}$ summarizing SU(3)--flavour breaking effects.
Note that $m_t$ and $V_{cb}$ dependences have been eliminated this way
 and $R_{ds}$
contains much smaller theoretical
uncertainties than the hadronic matrix elements in $x_d$ and $x_s$
separately.
Provided $x_d/x_s$ has been accurately measured a determination
of $R_t$ within $\pm 10\%$ should be possible.
The main features of $x_d/x_s$ are:
\begin{itemize}
\item No $\Lambda_{\overline{MS}}$, $m_t$ and $V_{cb}$ dependence.
\item Hadronic uncertainty in SU(3)--flavour breaking effects of
      roughly $\pm 10\%$.
\end{itemize}
Because of the last feature, $x_d/x_s$ cannot fully compete in the clean
determination of CKM parameters with CP asymmetries in B-decays and
with $\klpnn$. Although $\kpnn$ has smaller hadronic uncertainties
than $x_d/x_s$, its dependence on $\Lambda_{\overline{MS}}$ and  $m_c$
puts it in the same class as $x_d/x_s$ \cite{AB94A}.
\section{$\sin (2\beta)$ from $K\to \pi\nu\bar\nu$}
It has been pointed out in \cite{BH} that
measurements of $Br(\kpn)$ and $Br(\klpn)$ could determine the
unitarity triangle completely provided $m_t$ and $V_{cb}$ are known.
In view of the strong dependence of these branching ratios on
$m_t$ and $V_{cb}$ this determination is not precise however \cite{BB4}.
On the other hand it has been noticed recently \cite{BB4} that the $m_t$ and
$V_{cb}$ dependences drop out in the evaluation of $\sin(2\beta)$.
Introducing the "reduced" branching ratios
\begin{equation}\label{19}
B_+={Br(\kpn)\o 4.64\cdot 10^{-11}}\qquad
B_L={Br(\klpn)\o 1.94\cdot 10^{-10}}
\end{equation}
one finds
\begin{equation}\label{22}
\sin (2\beta)=\frac{2 r_s(B_+, B_L)}{1+r^2_s(B_+, B_L)}
\end{equation}
where
\begin{equation}\label{21}
r_s(B_+, B_L)=
{\sqrt{(B_+-B_L)}-P_0(K^+)\o\sqrt{B_L}}
\end{equation}
so that $\sin (2\beta)$
does not depend on $m_t$ and $V_{cb}$.
Here $P_0(K^+)=0.40\pm0.09$ \cite{BB3,BB4} is a function of $m_c$ and
 $\Lambda_{\overline{MS}}$ and includes the residual uncertainty
due to the renormalization scale $\mu$.
Consequently $\kpnn$ and $\klpnn$ offer a
clean
determination of $\sin (2\beta)$ which can be confronted with
the one possible in $B^0\to\psi K_S$ discussed above.
Any difference in these two determinations
would signal new physics.
Choosing
$Br(\kpn)=(1.0\pm 0.1)\cdot 10^{-10}$ and
$Br(\klpn)=(2.5\pm 0.25)\cdot 10^{-11}$,
one finds \cite{BB4}
\begin{equation}\label{26}
\sin(2 \beta)=0.60\pm 0.06 \pm 0.03 \pm 0.02
\end{equation}
where the first error is "experimental", the second represents the
uncertainty in $m_c$ and  $\Lambda_{\overline{MS}}$ and the last
is due to the residual renormalization scale uncertainties. This
determination of $\sin(2\beta)$ is competitive with the one
expected at the B-factories at the beginning of the next decade.
\section{Precise Determinations of the CKM Matrix}
Using the first two finalists and $\lambda=0.2205\pm 0.0018$
\cite{LR}
it is possible to determine all the parameters
of the CKM matrix without any hadronic uncertainties
\cite{AB94}.
With $a\equiv \sin(2\alpha)$, $b\equiv \sin(2\beta)$
and $Br(K_L)\equiv Br(\klpnn)$
one determines $\varrho$, $\eta$
and $\vcb$ as follows \cite{AB94}:
\begin{equation}\label{5}
\bar\varrho = 1-\bar\eta r_{+}(b)\quad ,\quad
\bar\eta=\frac{r_{-}(a)+r_{+}(b)}{1+r_{+}^2(b)}
\end{equation}
\begin{equation}\label{9b}
\mid V_{cb}\mid=0.039\sqrt\frac{0.39}{\eta}
\left[ \frac{170 ~GeV}{m_t} \right]^{0.575}
\left[ \frac{Br(K_L)}{3\cdot 10^{-11}} \right]^{1/4}
\end{equation}
where
\begin{equation}\label{7}
r_{\pm}(z)=\frac{1}{z}(1\pm\sqrt{1-z^2})
\qquad
z=a,b
\end{equation}
We note that the weak dependence of $\mid V_{cb}\mid$ on $Br(\klpnn)$
allows to achieve high accuracy for this CKM element even when $Br(\klpnn)$
is not measured precisely.

As illustrative examples we consider in table 1 three scenarios.
The first four rows give the assumed input parameters and their
experimental errors. The remaining rows give the results for
selected parameters. Further results can be found in \cite{AB94}.
The accuracy in the scenario I should be achieved at B-factories,
HERA-B,
at KAMI and at KEK.
 Scenarios II and
III correspond to B-physics at Fermilab during the Main Injector
era and at LHC
respectively.
At that time an improved measurement of $Br(\klpnn)$ should be aimed for.
\begin{table}
\begin{center}
\begin{tabular}{|c||c||c|c|c|}\hline
& Central &$I$&$II$&$III$\\ \hline
$\sin(2\alpha)$ & $0.40$ &$\pm 0.08$ &$\pm 0.04$ & $\pm 0.02 $\\ \hline
$\sin(2\beta)$ & $0.70$ &$\pm 0.06$ &$\pm 0.02$ & $\pm 0.01 $\\ \hline
$m_t$ & $170$ &$\pm 5$ &$\pm 3$ & $\pm 3 $\\ \hline
$10^{11} Br(K_L)$ & $3$ &$\pm 0.30$ &$\pm 0.15$ & $\pm 0.15 $\\ \hline\hline
$\varrho$ &$0.072$ &$\pm 0.040$&$\pm 0.016$ &$\pm 0.008$\\ \hline
$\eta$ &$0.389$ &$\pm 0.044$ &$\pm 0.016$&$\pm 0.008$ \\ \hline
$\mid V_{ub}/V_{cb}\mid$ &$0.087$ &$\pm 0.010$ &$\pm 0.003$&$\pm 0.002$
 \\ \hline
$\mid V_{cb}\mid/10^{-3}$ &$39.2$ &$\pm 3.9$ &$\pm 1.7$&$\pm 1.3$\\ \hline
$\mid V_{td}\mid/10^{-3}$ &$8.7$ &$\pm 0.9$ &$\pm 0.4$ &$\pm 0.3$ \\
 \hline \hline
$\mid V_{cb}\mid/10^{-3}$ &$41.2$ &$\pm 4.3$ &$\pm 3.0$&$\pm 2.8$\\ \hline
$\mid V_{td}\mid/10^{-3}$ &$9.1$ &$\pm 0.9$ &$\pm 0.6$ &$\pm 0.6$ \\
 \hline
\end{tabular}
\end{center}
\centerline{}
\caption{Determinations of various parameters in scenarios I-III }
\end{table}
Table 1 shows very clearly the potential of CP asymmetries
in B-decays and of $\klpnn$ in the determination of CKM parameters.
It should be stressed that this high accuracy is not only achieved
because of our assumptions about future experimental errors in the
scenarios considered, but also because $\sin(2\alpha)$ is a
very sensitive function of $\varrho$ and $\eta$ \cite{BLO},
$Br(\klpnn)$
depends strongly on $\mid V_{cb}\mid$ and most importantly because
of the clean character of the quantities considered.

It is instructive to investigate whether the use of
$\kpnn$ instead of $\klpnn$ would also give interesting results for
$V_{cb}$ and $V_{td}$.
 We again consider scenarios I-III with
$Br(\kpnn)= (1.0\pm 0.1)\cdot 10^{-10}$ for the scenario I and
$Br(\kpnn)= (1.0\pm 0.05)\cdot 10^{-10}$ for scenarios II and III
in place of $Br(\klpnn)$ with all other input parameters unchanged.
An analytic formula for $\vcb$ can be found in \cite{AB94}.
The results for $\varrho$, $\eta$, and $\mid V_{ub}/V_{cb}\mid$
remain of course unchanged. In the last two rows of table 1 we show the
results for $\mid V_{cb} \mid$ and $\mid V_{td}\mid$ .
We observe that
due to the uncertainties present in the charm contribution to
$\kpnn$, which was absent in $\klpnn$, the determinations of
 $\mid V_{cb}\mid$ and  $\mid V_{td}\mid$  are less accurate.
 If the uncertainties due to the charm mass
and $\Lambda_{\overline{MS}}$ are removed one day this analysis
will be improved \cite{AB94}.

An alternative strategy is to use the measured value of $R_t$ instead
of $\sin(2\alpha)$. Then (\ref{5})
is replaced by
\begin{equation}\label{5a}
\bar\varrho = 1-\bar\eta r_{+}(b)\quad ,\quad
\bar\eta=\frac{R_t}{\sqrt{2}}\sqrt{b r_{-}(b)}
\end{equation}
The result of this exercise is shown in table 2. We observe that even
with rather optimistic assumptions on the accuracy of
$R_t$, this determination of
CKM parameters cannot fully compete with the previous one.
Again the last two rows give the results when $\klpnn$ is replaced
by $\kpnn$.
\begin{table}
\begin{center}
\begin{tabular}{|c||c||c|c|c|}\hline
& Central &$I$&$II$&$III$\\ \hline
$R_t$ & $1.00$ &$\pm 0.10$ &$\pm 0.05$ & $\pm 0.03 $\\ \hline
$\sin(2\beta)$ & $0.70$ &$\pm 0.06$ &$\pm 0.02$ & $\pm 0.01 $\\ \hline
$m_t$ & $170$ &$\pm 5$ &$\pm 3$ & $\pm 3 $\\ \hline
$10^{11} Br(K_L)$ & $3$ &$\pm 0.30$ &$\pm 0.15$ & $\pm 0.15 $\\ \hline\hline
$\varrho$ &$0.076$ &$\pm 0.111$&$\pm 0.053$ &$\pm 0.031$\\ \hline
$\eta$ &$0.388$ &$\pm 0.079$ &$\pm 0.033$&$\pm 0.019$ \\ \hline
$\mid V_{ub}/V_{cb}\mid$ &$0.087$ &$\pm 0.014$ &$\pm 0.005$&$\pm 0.003$
 \\ \hline
$\mid V_{cb}\mid/10^{-3}$ &$39.3$ &$\pm 5.7$ &$\pm 2.6$&$\pm 1.8$\\ \hline
$\mid V_{td}\mid/10^{-3}$ &$8.7$ &$\pm 1.2$ &$\pm 0.6$ &$\pm 0.4$ \\
 \hline \hline
$\mid V_{cb}\mid/10^{-3}$ &$41.3$ &$\pm 5.8$ &$\pm 3.7$&$\pm 3.3$\\ \hline
$\mid V_{td}\mid/10^{-3}$ &$9.1$ &$\pm 1.3$ &$\pm 0.8$ &$\pm 0.7$ \\
 \hline
\end{tabular}
\end{center}
\centerline{}
\caption{ As in table 1 but with $\sin(2\alpha)$ replaced by $R_t$.}
\end{table}
\section{Final Remarks}
\begin{itemize}
\item Precise measurements of all CKM parameters without hadronic
uncertainties are possible.
\item Such measurements are essential for the tests of the
standard model.
Of particular interest will be the comparison of $\mid V_{cb}\mid$
determined as suggested here with the value of this CKM element extracted
from tree level semi-leptonic  B-decays.
 Since in contrast to
$\klpnn$ and $\kpnn$, the tree-level decays are to an excellent approximation
insensitive to any new physics contributions from very high energy scales,
the comparison of these two determinations of $\mid V_{cb}\mid$ would
be a good test of the standard model and of a possible physics
beyond it.
\end{itemize}
Precise determinations of all CKM parameters without hadronic uncertainties
 along the lines presented
here can only be realized if the measurements of CP asymmetries in
B-decays and the measurements of $Br(\klpnn)$, $Br(\kpnn)$ and $x_d/x_s$
can reach the desired accuracy.
All efforts should be made to achieve this goal.

\end{document}